\documentclass[aps,pre,showpacs,twocolumn]{revtex4}

\usepackage{graphicx}
\usepackage{float}
\usepackage{color}
\usepackage{amsmath}
\usepackage{amsfonts}
\usepackage{amssymb}
\usepackage{wrapfig}
\usepackage{url}
\usepackage{bbold}

\begin{document}

\title{Tensor-network study of quantum phase transition on Sierpi\'nski fractal}

\author{Roman Krcmar$^1$, Jozef Genzor$^2$, Yoju Lee$^3$, Hana \v{C}en\v{c}arikov\'{a}$^4$, 
Tomotoshi Nishino$^2$, and Andrej Gendiar$^1$}
\affiliation{$^1$Institute of Physics, Slovak Academy of Sciences, SK-845 11, Bratislava, Slovakia}
\affiliation{$^2$Department of Physics, Graduate School of Science, Kobe University, Kobe 657-8501, Japan}
\affiliation{$^3$Faculty of Physics, University of Vienna, Boltzmanngasse 5, A-1090 Vienna, Austria}
\affiliation{$^4$Institute of Experimental Physics, Slovak Academy of Sciences, SK-040 01 Ko\v{s}ice, Slovakia}

\begin{abstract}
The transverse-field Ising model on the Sierpi\'nski fractal, which is characterized by the fractal dimension 
$\log_2^{~} 3 \approx 1.585$, is studied by a tensor-network method, the Higher-Order Tensor Renormalization
Group. We analyze the ground-state energy and the spontaneous magnetization in the thermodynamic limit. 
The system exhibits the second-order phase transition at the critical transverse field $h_{\rm c}^{~} = 1.865$. 
The critical exponents $\beta \approx 0.198$ and $\delta \approx 8.7$ are obtained. Complementary to the 
tensor-network method, we make use of the real-space renormalization group and improved mean-field 
approximations for comparison.
\end{abstract}

\maketitle

\section{Introduction}

The classification of quantum phase transitions remains one of the major interests in the condensed matter 
physics. Although there are groups of exactly solvable models in physics, a vast majority of the physical systems 
calls for different approaches, in particular, for numerical calculations. Some of them are straightforwardly applicable,
such as the Monte Carlo (MC) simulations, whereas the other ones, including renormalization group techniques,
require development of novel algorithms.

This work is oriented for classification of the quantum phase transition on a fractal lattice, which is the infinite-size 
Sierpi\'nski fractal (triangle or gasket), whose Hausdorff fractal dimension is $\log_2^{~} 3 \approx 1.585$. 
It is known that the classical Ising model exhibits no phase transition on the Sierpi\'nski fractal~\cite{gefen,desai}. 
Substantially less is known about its quantum counterpart. A couple of recent works investigated quantum 
spin models on fractals by means of real-space renormalization-group methods and by classical MC 
simulations~\cite{Yi,kubica1,kubica2,wang,xu}. 

In order to bring more light into the quantum fractal system, we consider 
a different methodology, which is the Higher-Order Tensor Renormalization Group (HOTRG) method~\cite{hotrg}.
It should be noted that the tensor-network viewpoint is efficient for expressing the recursive
structure of the fractal lattices~\cite{jozofraktal}. 
In particular, we focus on the quantum Ising model on the Sierpi\'nski fractal, and analyze
the ground-state energy per site $E_0^{~}$ and the spontaneous magnetization 
$\langle \sigma^z_{~} \rangle$ with respect to the transverse field $h_x^{~}$.
We first determine the critical field $h_{\rm c}^{~}$, and then we estimate the critical 
exponents $\beta$ and $\delta$ from the calculated $\langle \sigma^z_{~} \rangle$. 

Structure of this article is as follows. In the next Section, we explain the lattice structure,
and introduce the system Hamiltonian. We first consider two conventional calculation methods,
one is the improved mean-field approximations, and the other is the real-space renormalization-group
(RSRG) method.  The way of applying the HOTRG method for this fractal system is
presented in Sec.~III. We show the numerical result in Sec.~IV. Conclusions are summarized in the last section. 
In the Appendix, we discuss the numerical stabilization in the HOTRG method, 
which is realized with the correct initialization of the tensor. 
Two types of entanglement entropies, the vertical and the horizontal ones, of the local tensor 
are compared, since their ratio quantifies the anisotropy in the tensor. 
When they are comparable, one can avoid the instability.

\section{model and conventional approximations}\label{sec:met}

\begin{figure}[tb]
\includegraphics[width = 0.45\textwidth,clip]{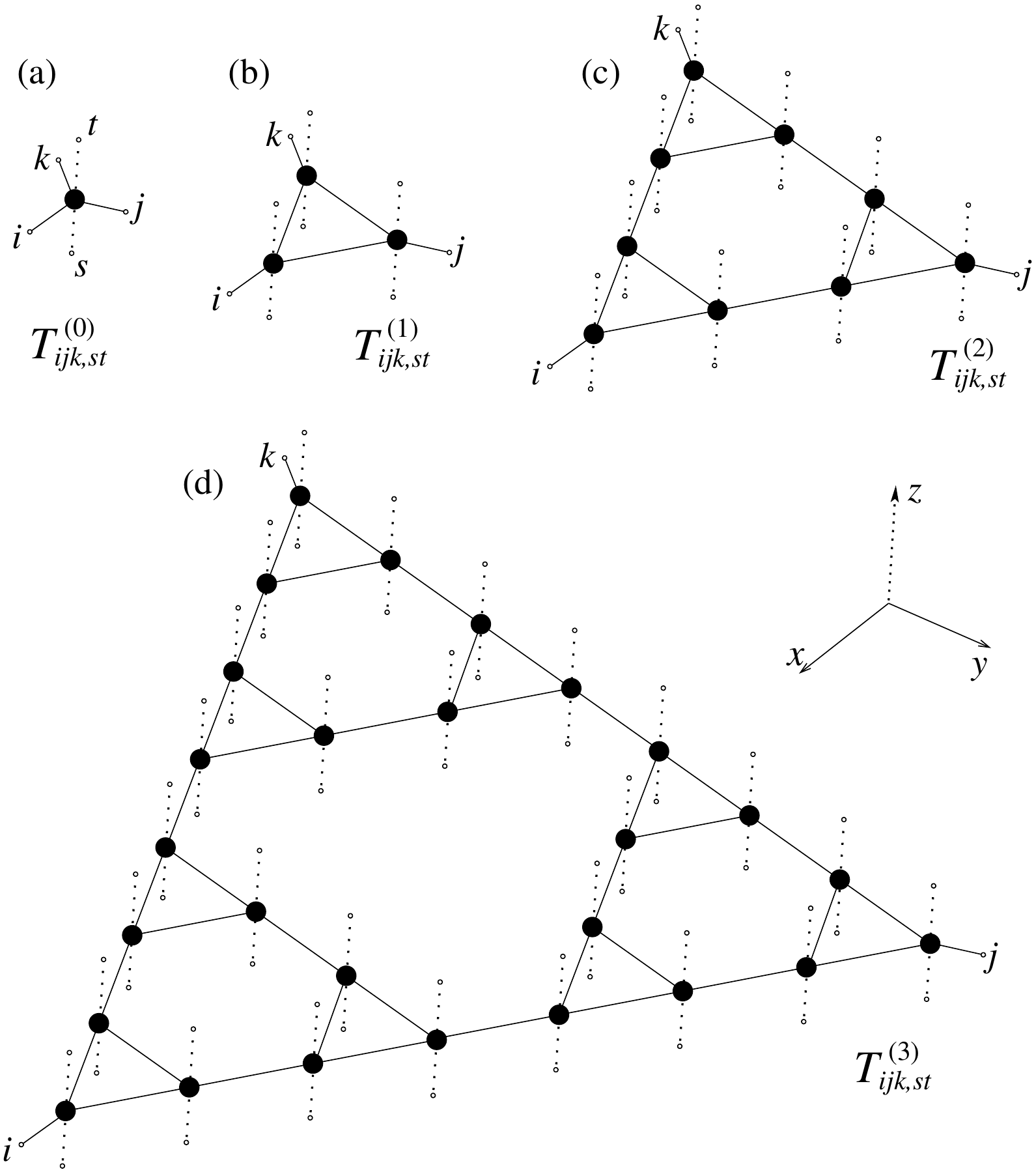}
\caption{Structure of the Sierpi\'nski fractal. (a) The smallest unit consists of a site shown by the black circle,
from which the three bonds $i$, $j$, and $k$ emerge. (b)-(d) Connecting the three units, one can iteratively
expand the size of the unit. The vertical dotted lines correspond to the imaginary-time evolution,
which is considered in Sec.~III.}
\label{fig:tensor}
\end{figure}

Figure~\ref{fig:tensor} shows the structure of the Sierpi\'nski fractal.
The lattice is recursively constructed by connecting three units, as shown in (a)-(d). 
The black dots represent the lattice sites, and the full lines represent the nearest-neighboring 
connections, the bonds. The $x$- and $y$-axes are used to denote the two-dimensional plane on
which the fractal is located. The $x$-axis is parallel to the bonds 
denoted by $i$, whereas the $y$-axis is not parallel to $j$. The $z$-axis
is perpendicular to the plane. 
For the moment, let us omit the vertical (dotted) lines and disregard the tensor notations shown in the figure.

The Hamiltonian of the transverse-field Ising model on the lattice has the form
\begin{equation}
H = - J \sum_{\langle a,b \rangle}^{~} \sigma^z_a \sigma^z_b
   - h_x^{~} \sum_{a}^{~} \sigma^x_a
   - h_z^{~} \sum_{a}^{~} \sigma^z_a \, ,
\label{HmTFIM}
\end{equation}
where $\sigma^x_a$ and $\sigma^z_a$ represent the Pauli spin operators acting on the lattice site $a$. The uniform
fields $h_x^{~}$ and $h_z^{~}$, respectively, are applied to the transverse ($x$-) and longitudinal 
($z$-) directions. The ferromagnetic Ising interaction $J > 0$ is present between the nearest-neighboring spin
pairs $\sigma^z_a$ and $\sigma^z_b$. The interacting pairs are denoted by the symbol $\langle a, b \rangle$, 
and they are located on bonds in the fractal lattice. Throughout this article we focus on the ground-state of this
system and its quantum phase transition with respect to the transverse field $h_x^{~}$. Hereafter we assume $J = 1$.
The parameter $h_z^{~}$ is set to zero unless its value is specified.

Before we start explaining the details of the HOTRG method, we briefly introduce the two conventional approximation
schemes. The first one is the mean-field approximation, which offers a rough insight into the ground-state. We consider
three types of the gradually improving mean-field approximations, beginning from the smallest unit size shown in Fig.~1(a),
followed by an extended unit size with the 3 sites in Fig.~1(b), and finally that with the 9 sites in Fig.~1(c).
All of the interactions inside each extended unit are treated rigorously while the inter-unit Ising interaction
through the bonds $i$, $j$, and $k$ are replaced by the mean-field value $-J \langle \sigma_{~}^{z} \rangle$.
Here, the average of the bond energy and magnetization is taken over all the sites in the extended unit.
Thermodynamic functions could be estimated with a better precision if the size of the unit gets larger.
We confirm this conjecture in Sec. IV.

The second approximation scheme we consider is the conventional real-space renormalization group (RSRG) method~\cite{rg}, 
which shares some aspects in common with the HOTRG method. The RSRG method consists of an iterative procedure, where
the effective intra-unit Hamiltonian $H^{(\ell)}_{a}$ for $\ell=0,1,2,\dots$ is created recursively.
(The expanded unit contains $3^{\ell}_{~}$ sites.) Since we intend to estimate the critical field $h_{\rm c}^{~}$ only,
we explain the case when $h_z^{~} = 0$.
At the initial step, which corresponds to the single site in Fig.~1(a), we have $H^{(0)}_{a} = - h_x^{~} \,
\sigma_{a}^x$ and the spin operators $\sigma_{a;i}^{(0)} = \sigma_{a;j}^{(0)} =
\sigma_{a;k}^{(0)} = \sigma_{a}^z$, where $a$ specifies the site location, and $i$, $j$, $k$
denote the pairing directions of the neighboring interactions.

Let us consider the 3-site extended unit shown in Fig.~1(b), and label the sites as $a$ (left), $b$ (right),
and $c$ (top). The Hamiltonian of this 3-site unit is then written as
\begin{eqnarray}
{\cal H}^{(\ell)}_{~} &=& H_{a}^{(\ell)} + H_{b}^{(\ell)} + H_{c}^{(\ell)} \nonumber\\
&-& 
\sigma^{(\ell)}_{a;j} \sigma^{(\ell)}_{b;i} - 
\sigma^{(\ell)}_{b;k} \sigma^{(\ell)}_{c;j} - 
\sigma^{(\ell)}_{c;i} \sigma^{(\ell)}_{a;k} 
	\label{rg1}
\end{eqnarray}
at the initial iteration step $\ell = 0$. After diagonalizing ${\cal H}^{(\ell)}_{~}$, we keep only those eigenstates
that are associated with $D$ lowest eigenvalues (while the remaining high-energy eigenstates are discarded).
The renormalization group transformation $U$ is then chosen to the projection 
to the low-energy eigenstates, which reduce the dimension down to $D$. 
Applying $U$ to ${\cal H}^{(\ell)}_{~}$, we obtain the renormalized
intra-unit Hamiltonian 
\begin{equation}
H^{(\ell+1)}_{\chi} = U^{\dagger}_{~}  {\cal H}^{(\ell)}_{~} \, U_{~}
\end{equation}
for the extended unit labeled by $\ell + 1$, where $\chi=a,b,c$ is the site index for the extended unit.
In the same manner, we obtain the renormalized $z$-component of the spin 
\begin{eqnarray}
\sigma^{(\ell+1)}_{a;i} &=& U^{\dagger}_{~} \sigma^{(\ell)}_{a;i} \, U_{~} , \nonumber\\
\sigma^{(\ell+1)}_{b;j} &=& U^{\dagger}_{~} \sigma^{(\ell)}_{b;j} \, U_{~} , \nonumber\\
\sigma^{(\ell+1)}_{c;k} &=& U^{\dagger}_{~} \sigma^{(\ell)}_{c;k} \, U_{~} , 
\end{eqnarray}
at each corner of the extended unit. At this point, we can return to Eq.~\eqref{rg1} to obtain an effective 
Hamiltonian ${\cal H}^{(\ell+1)}_{~}$ for the 9-site unit shown in Fig.~1(c). As we show in Sec. IV, the transition
point (the critical phase transition field) is obtained with relatively high numerical precision if a sufficiently
large $D$, the number of the block-spin state, is taken.

\section{Higher-Order Tensor Renormalization Group}

We focus on the numerical analysis of the quantum fractal system by means of the 
HOTRG method~\cite{hotrg}, which has yielded a high numerical accuracy for two- and three-dimensional 
classical Ising model. The method has also been applied to one- and two-dimensional quantum Ising
model through the quantum-classical correspondence, which is a discrete imaginary-time 
path-integral representation. The imaginary-time evolution expressed by the density operator 
$\rho = e^{- \tau {\cal H}}_{~}$ is essential, as it behaves as the projection to the ground-state in 
the large $\tau$ limit. 

\begin{figure*}[tb]
\includegraphics[width=0.98\textwidth,clip]{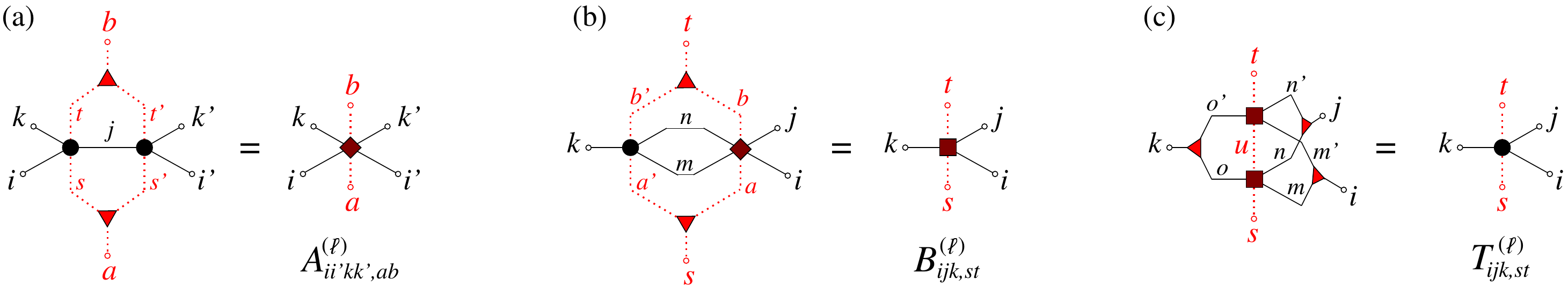}
\caption{(Color online) Graphical representation of the processes in the HOTRG method in 
Eqs.~\eqref{HOTRGa}-\eqref{HOTRGc}. We use the large symbols to indicate the tensors 
$T^{(\ell)}_{~}$ (circles), $A^{(\ell)}_{~}$ (diamonds), $B^{(\ell)}_{~}$ (squares), and the projectors 
$U$, $U'$, and $U''$ (triangles).}
\label{fig:Gtrg}
\end{figure*}

Let us consider the Hamiltonian in Eq.~\eqref{HmTFIM} and divide the imaginary-time span 
$\tau$ into $m$ intervals $\Delta\tau = \tau / m$. We express $\rho$ in the form of product 
\begin{equation}
	\rho = \left( e^{- \Delta\tau {\cal H}}_{~} \right)^{m}_{~} = 
	\left[ e^{-\Delta\tau \left( {\cal H}_{zz}^{~} + {\cal H}_{x}^{~} + {\cal H}_{z}^{~} \right)}_{~}
	\right]^{m}_{~} 
\end{equation}
among imaginary-time intervals, where ${\cal H}_{zz}^{~}$, ${\cal H}_{x}^{~}$, and ${\cal H}_{z}^{~}$,
respectively, correspond to the first, second, and third term in the r.h.s. of Eq.~\eqref{HmTFIM}. Although
${\cal H}_{x}^{~}$ does not commute with ${\cal H}_{zz}^{~}$ or ${\cal H}_{z}^{~}$, a good approximation of $\rho$
can be obtained by means of the Trotter-Suzuki decomposition~\cite{trotter}
\begin{equation}
\rho \approx \left[
	e^{-\Delta\tau \left( {\cal H}_{zz}^{~} + {\cal H}_{z}^{~} \right)}_{~} \, 
	e^{-\Delta\tau {\cal H}_{x}^{~}}_{~}
\right]^{m}_{~}\, ,
\label{TSD}
\end{equation}
provided that $\Delta\tau$ is sufficiently small (e.g. $\Delta\tau\approx0.01$). 
Each imaginary-time interval, which corresponds to $e^{- \Delta\tau {\cal H}}_{~}$, plays the role
of the transfer matrix.

Applying a duality transformation, which introduces new two-state variables in the middle of the connected bonds,
we can express the transfer matrix in terms of the tensor network. Figure~\ref{fig:tensor} shows the structure of the 
transfer matrix for the elementary unit, in (a) and for the extended ones, in (b)-(d). This time, we regard the black
dots as local tensors, which have three legs in the spatial direction shown by the lines, and two legs in the
imaginary-time directions shown by the vertical dotted lines. Each local tensor is given by
\begin{equation}
T^{(0)}_{ijk,st} = \sum_{\sigma}^{~}
W_{\sigma i}^{~}
W_{\sigma j}^{~}
W_{\sigma k}^{~}
P_{\sigma s}^{~}
P_{\sigma t}^{~}
G_{\sigma}^{~} \, ,
\label{iniT}
\end{equation}
where the matrix
\begin{equation}
W = \left(
\begin{array}{rr}
\sqrt{ \cosh(J \Delta\tau)} &  \sqrt{\sinh(J \Delta\tau) }\\[0.5mm]
\sqrt{ \cosh(J \Delta\tau)} & -\sqrt{\sinh(J \Delta\tau) }
\end{array}
\right)
\end{equation}
originates from the Ising interaction in ${\cal H}_{zz}^{~}$ between the neighboring spins. The other matrix
\begin{equation}
P = \frac{1}{\sqrt{2}}\left(
\begin{array}{rr}
\exp( h_x^{~} \Delta\tau/2) &  \exp(-h_x^{~} \Delta\tau/2)\\[0.5mm]
\exp( h_x^{~} \Delta\tau/2) & -\exp(-h_x^{~} \Delta\tau/2)
\end{array}
\right)
\end{equation}
corresponds to the spin-flipping effect by the transverse field $h_x^{~}$ in ${\cal H}_x^{~}$, and the column vector
\begin{equation}
G = \left(
\begin{array}{c}
\exp(\phantom{+}h_z^{~} \Delta\tau)\\[0.5mm]
\exp(          -h_z^{~} \Delta\tau)
\end{array}
\right)
\end{equation}
represents the effect of external field $h_z^{~}$ along the z-direction in ${\cal H}_z^{~}$. 
All the indices $i$, $j$, $k$, $s$, and $t$ of $T^{(0)}_{ijk,st}$ thus carry two degrees of the freedom. 

The transfer matrices 
$T^{(1)}_{ijk,st}$,  
$T^{(2)}_{ijk,st}$, and
$T^{(3)}_{ijk,st}$ for the extended units, respectively, shown in Fig.~\ref{fig:tensor}(b), (c), and (d),
can be obtained by contracting horizontal legs in a recursive manner. Actually, we do not directly treat
these extended transfer matrices. Our aim is to obtain the local thermodynamic quantities
\begin{equation}
\langle O \rangle = 
	\frac{ {\rm Tr} \, \left( \, O \rho \, \right) }{ {\rm Tr} ~ (\rho) } = 
\frac{ {\rm Tr} \, \left( \, O \, 
	e^{-\tau {\cal H}}_{~} \, \right) }
	{ {\rm Tr} ~ \left( e^{-\tau {\cal H}}_{~} \right) } \, ,
\label{SO}
\end{equation}
where $O$ represent a local operator, for a sufficiently wide system when $\tau$ is large enough. 
For this purpose, we do not have to construct $\rho$ in a faithful manner, but only need to consider
a series of finite-size clusters represented as a stack of the extended transfer matrices, i.e.,
\begin{equation*}
\left[ T^{(1)}_{ijk,st} \right]^2, 
\left[ T^{(2)}_{ijk,st} \right]^4, 
\left[ T^{(3)}_{ijk,st} \right]^8,
\dots,
\left[ T^{(\ell)}_{ijk,st} \right]^{2^{\ell}},
\dots
\end{equation*}
By use of these tensors and another series of tensors that contain the operator $O$ inside, 
the following ratio
\begin{equation}
\lim_{\tau \rightarrow \infty}^{~} \, 
\frac{\langle \phi | \, e^{-\tau {\cal H} / 2}_{~} \, O \, e^{-\tau {\cal H} / 2}_{~} | \phi \rangle}
{\langle \phi | \, e^{-\tau {\cal H} / 2}_{~} ~ e^{-\tau {\cal H} / 2}_{~} | \phi \rangle}
\end{equation}
can be obtained, which coincides with $\langle O \rangle$ in Eq.~(11) for a wide choice of
the boundary conditions and the trial state represented by $| \phi \rangle$.
The HOTRG method is appropriate for this purpose. Alternatively, the value in Eq.~(12) can
be obtained by use of tensor product state~\cite{tps1,tps2} and also the projected entangled pair 
state~\cite{peps}, but the computational cost is much higher. 

We create the stack of the transfer matrices in a renormalized form, through the recursive contraction processes, 
\begin{subequations}
\begin{align}
\label{HOTRGa}
&A_{ii'kk',ab}^{(\ell)} = \sum_{jss'tt'}^{~}
        T^{(\ell)}_{ijk,st} \, T^{(\ell)}_{ji'k',s't'} \, U_{ss',a}^{~} \, U_{tt',b}^{~} \, ,\\
\label{HOTRGb}
&B^{(\ell)}_{ijk,st}    = \sum_{\genfrac{}{}{0pt}{}{mn}{aa'bb'}}
        A_{ijmn,ab}^{(\ell)} \, T^{(\ell)}_{knm,a'b'} \, U'_{aa',s} \, U'_{bb',t} \, ,\\
\label{HOTRGc}
&T^{(\ell+1)}_{ijk,st}  = \sum_{\genfrac{}{}{0pt}{}{umm'}{nn'oo'}}
        B^{(\ell)}_{mno,su} B^{(\ell)}_{m'n'o',ut} U''_{mm',i} U''_{nn',j} U''_{oo',k} \, .
\end{align}
\end{subequations}
which are depicted by diagrams in Fig.~\ref{fig:Gtrg}. The projectors $U$, $U'$, and $U''$, which
are also called as isometries, are quasi unitary rectangular matrices of the size $D^2 \times D$, with $D$
being the degree of the freedom for a tensor index. These matrices are obtained from the higher-order 
singular value decomposition, whenever two tensors are combined and consequently reshaped into 
a matrix form~\cite{hotrg}. We keep the states that correspond to $D$ largest singular values.
Hence, the larger the $D$, the better the approximation is~\cite{jozofraktal,hosvd}.  The expansion
procedure is stopped after all of the thermodynamic functions (normalized per site) completely converge.

In this manner the HOTRG method, applied to the discrete path-integral representation of the quantum
fractal system, enables us to built up a sufficiently large finite-size system. Note that during the recursive
extension of the system, we can obtain thermodynamic functions such as the ground-state energy $E_0^{~}$ per 
site, as it is has been done for transverse-field Ising model on the square lattice. Further details on the
calculation of $E_0^{~}$ can be found in Refs.~\onlinecite{hotrg,jozofraktal}. One-point functions, such as
magnetization $\langle \sigma^z\rangle$, can also be calculated by introducing an impurity tensor, 
as discussed in Ref.~\onlinecite{jozofraktal}. In the appendix we discuss the choice of $\Delta \tau$
and the stabilization of numerical calculation by means of an appropriate choice of the initial tensor.

\section{Numerical results}

The mean-field approximation (MFA) offers a rough insight into phase transitions. As we have introduced in Sec.~II, 
we consider series of the three approximations, MFA$_1^{~}$, MFA$_3^{~}$, and MFA$_9^{~}$, respectively, where the 
interactions inside the (extended) units shown in Fig.~1 (a), (b), and (c) are treated exactly. We have also introduced
the RSRG method, which can capture critical behavior of the model, provided that a sufficiently large number of the
block-spin states $D$ is kept. However, the improvement in expectation values with respect to $D$ is rather slow.
In contrast, the numerical precision in the HOTRG method significantly improves with $D$.
We have confirmed that $D = 8$ is large enough to obtain well-converged results 
on this fractal lattice. We present the numerical results up to $D = 20$
in the HOTRG method.

\begin{figure}[tb]
\includegraphics[width = 0.48\textwidth,clip]{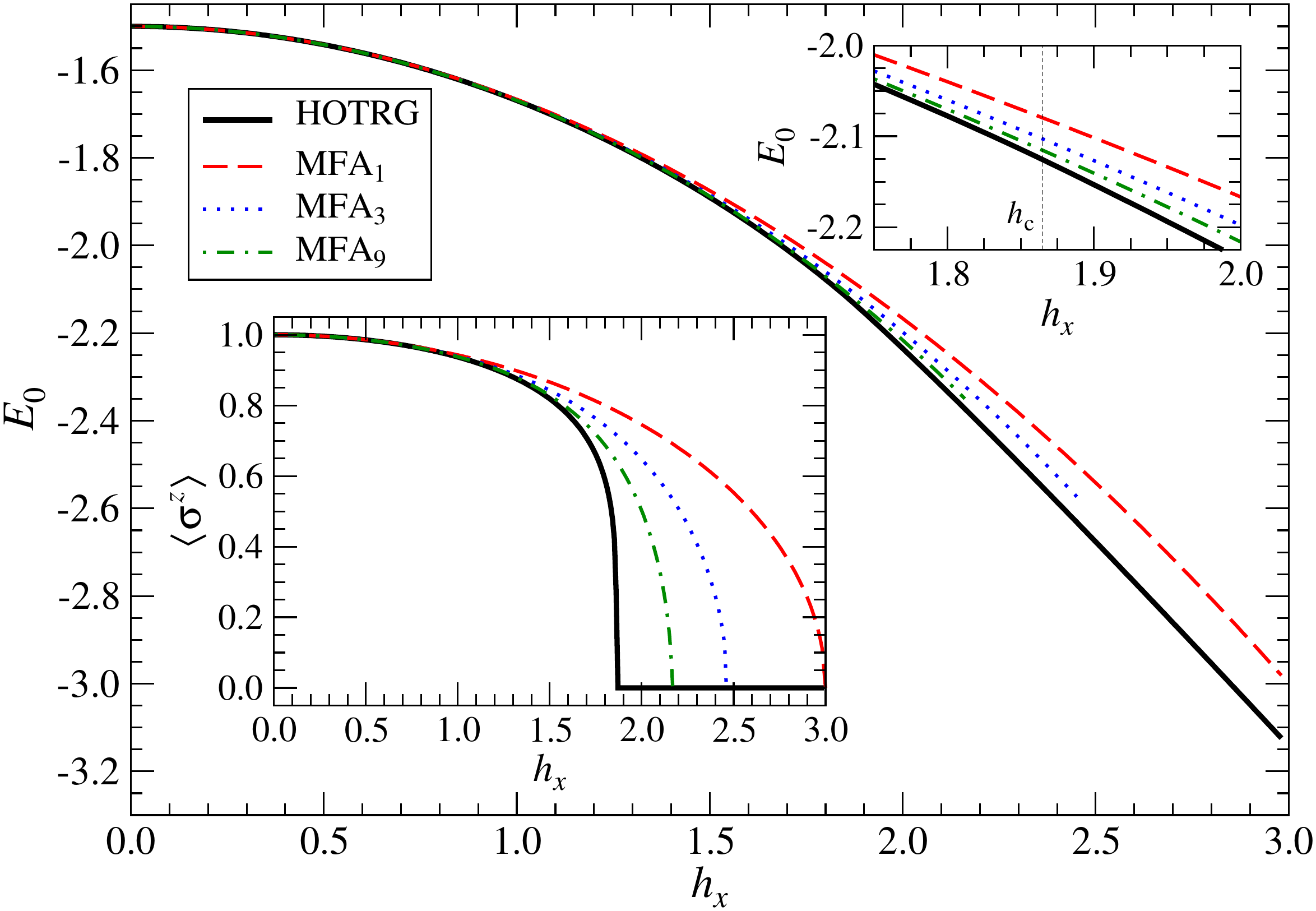}
\caption{(Color online) The ground-state energy per site $E_0^{~}$ versus 
transverse field $h_x^{~}$ at $h_z^{~} = 0$. The data obtained by the 
HOTRG method are shown by the thick full lines. The mean-field approximations MFA$_1^{~}$, 
MFA$_3^{~}$, and MFA$_9^{~}$ are, respectively, shown by dashed, dotted, and dot-dashed lines. 
The top inset shows $E_0^{~}$ around the transition point $h_{\rm c}^{~}$. The bottom inset shows 
$\langle \sigma^z_{~} \rangle$.}
\label{fig:HOTRG-MFAn}
\end{figure}

We first compare the three types of mean-field approximations with the HOTRG method when $D = 8$. The ground-state 
energy per site $E_0^{~}$ with respect to the transverse field $h_x^{~}$ is shown in Fig.~\ref{fig:HOTRG-MFAn}. 
The ground-state energy obtained by the HOTRG method is always the lower than those obtained by the mean-field
approximations. This can be better visible by comparing $E_0^{~}$ around the phase transition 
$h_x^{~} = h_{\rm c}^{~}$ as shown in the top inset. The bottom inset shows the spontaneous magnetization 
$\langle \sigma^z_{~} \rangle$. Since the fractal lattice is not homogeneous, the expectation value is
calculated by averaging three independent impurity operators~\cite{APS} contained in $T^{(1)}$,
cf Fig.~\ref{fig:tensor}(b).
From $\langle \sigma^z_{~} \rangle$ obtained by the HOTRG method, the critical field is determined as
$h_{\rm c}^{~} = 1.865$.

\begin{figure}[tb]
\includegraphics[width = 0.48\textwidth,clip]{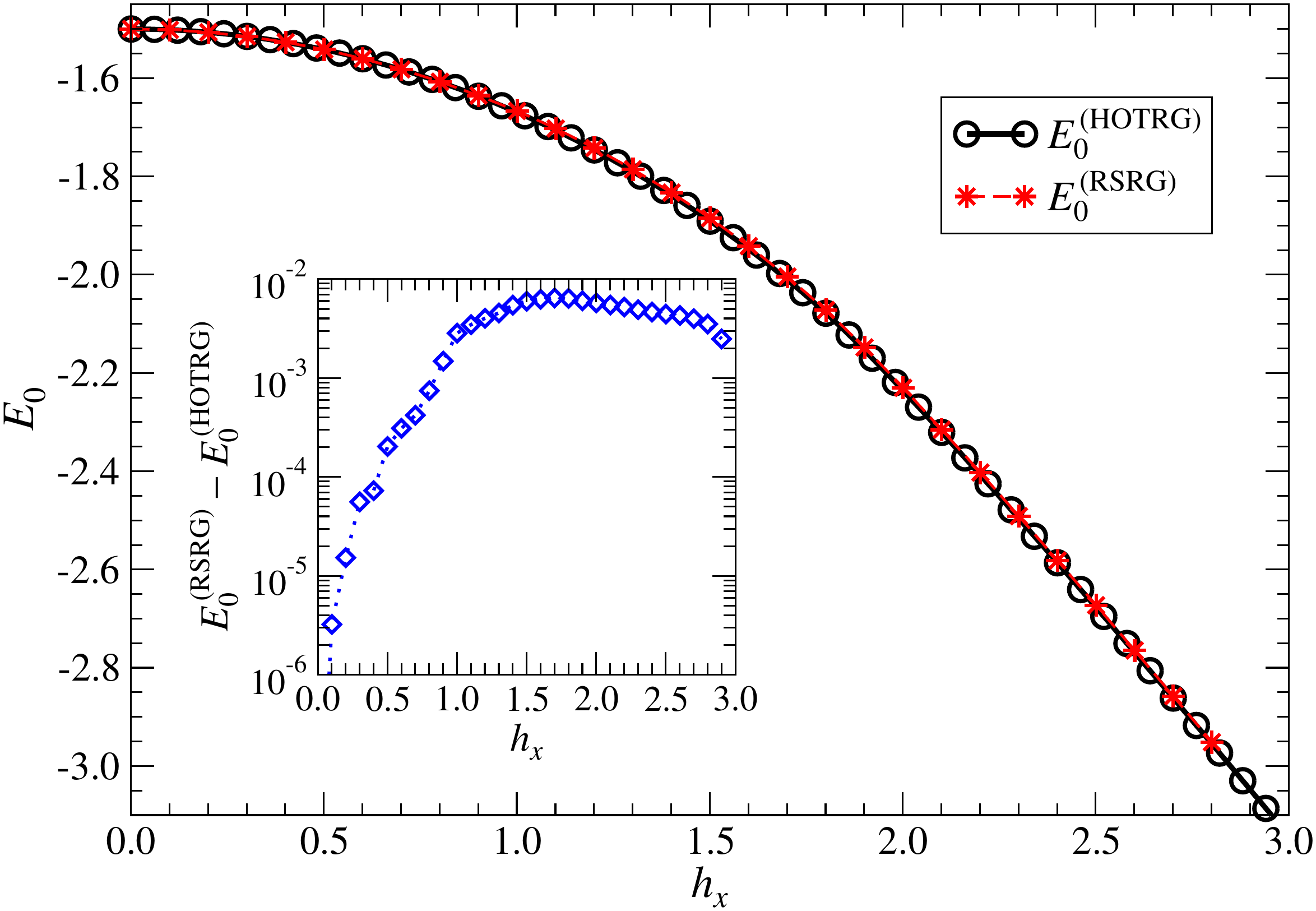}
\caption{(Color online) The ground-state energies per site with respect to $h_x^{~}$
calculated by the HOTRG method ($D = 8$) and by the RSRG method ($D = 24$) when $h_z^{~} = 0$. 
The inset shows their difference.}
\label{fig:energy}
\end{figure}
\begin{figure}[tb]
\includegraphics[width = 0.48\textwidth,clip]{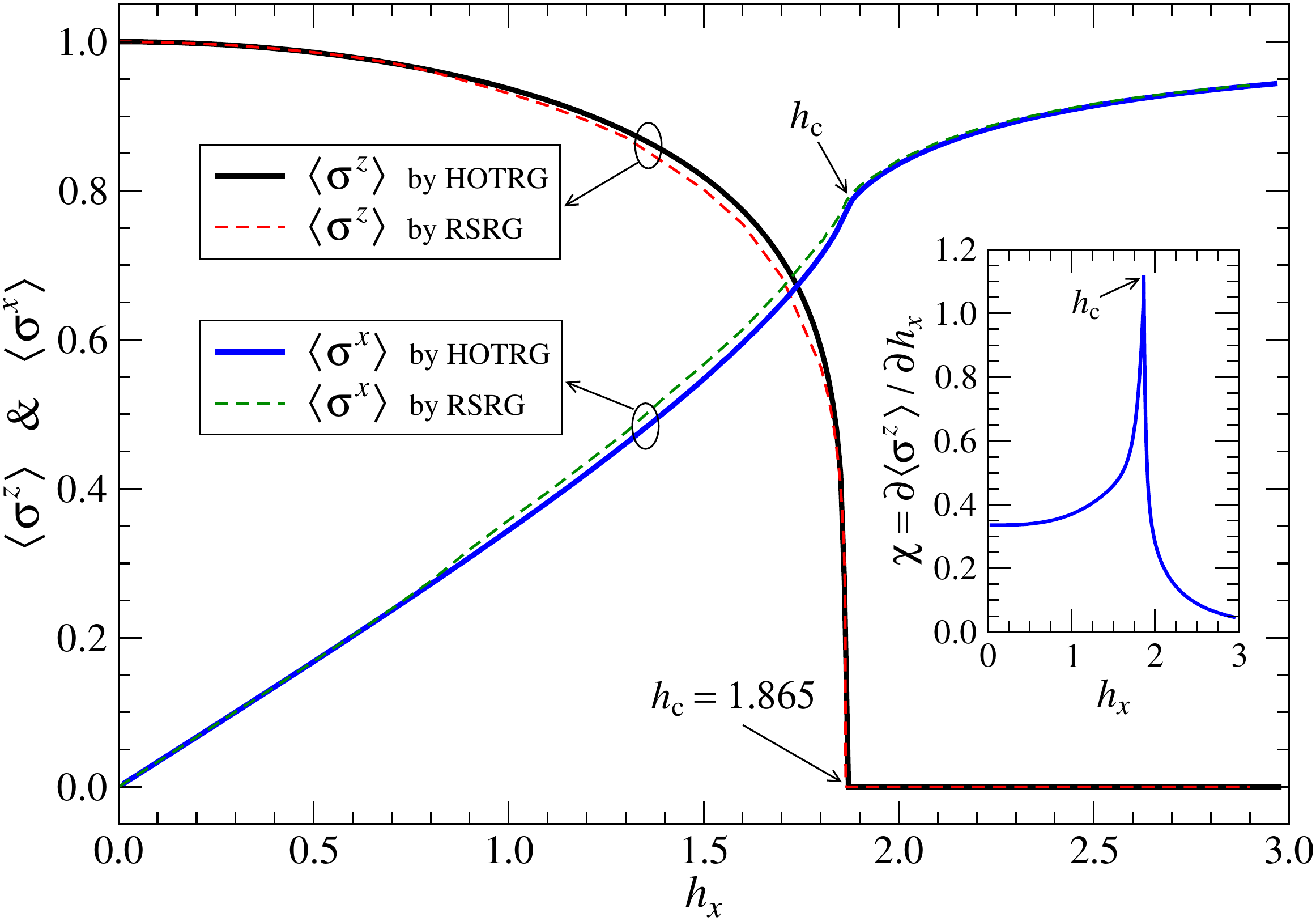}
\caption{(Color online) Spontaneous magnetization $\langle \sigma^z\rangle$ and the induced
polarization $\langle \sigma^x\rangle$ with respect to $h_x^{~}$. 
The full lines show the calculated result by the HOTRG method ($D = 8$) and the dashed ones
by the RSRG method ($D = 24$). The inset shows the susceptibility $\chi$ defined by Eq.~\eqref{magsusc}.}
\label{fig:mag}
\end{figure}

The RSRG method provides relatively accurate $E_0^{~}$ when $D = 24$ block-spin states are kept. 
Figure~\ref{fig:energy} shows $E_0^{~}$ obtained by the RSRG method ($D = 24$) 
and that by the HOTRG method ($D = 8$). The inset shows the difference in the calculated 
$E_0^{~}$, within the range $0 \leq h_x^{~} \leq 3$, which is not conspicuous. 
Figure~\ref{fig:mag} shows the spontaneous magnetization $\langle \sigma^z_{~} \rangle$
and induced polarization $\langle \sigma^x_{~} \rangle$. The difference between
both of the methods is better visible below the transition point. The 
spontaneous magnetization $\langle \sigma^z_{~} \rangle$ by the RSRG method gives
the critical field $h_{\rm c}^{~} = 1.864$, 
which is close to the value $h_{\rm c}^{~} =1.865$ determined by the HOTRG method.
The induced polarization $\langle \sigma^x_{~} \rangle$ is calculated by making use
of the Hellman-Feynman theorem~\cite{hellmann}
\begin{equation}
\langle \sigma^x_{~} \rangle = - \frac{\partial E_0}{\partial h_x}\, ,
\end{equation}
which exhibits a weak singularity at the critical field $h_x^{~} = h_{\rm c}^{~}$, as marked
by the arrow. (We still consider the case $h_z^{~} = 0$.) The inset of Fig.~\ref{fig:mag} shows
the susceptibility 
\begin{equation}
\chi = - \frac{\partial^2 E_0^{~}}{\partial h_x^2} = 
\frac{\partial \langle \sigma^x_{~} \rangle}{\partial h_x^{~}} \,
\label{magsusc}
\end{equation}
for the result of the HOTRG method, where there is a singular peak at $h_{\rm c}^{~}$. 

\begin{figure}[tb]
\includegraphics[width = 0.48\textwidth,clip]{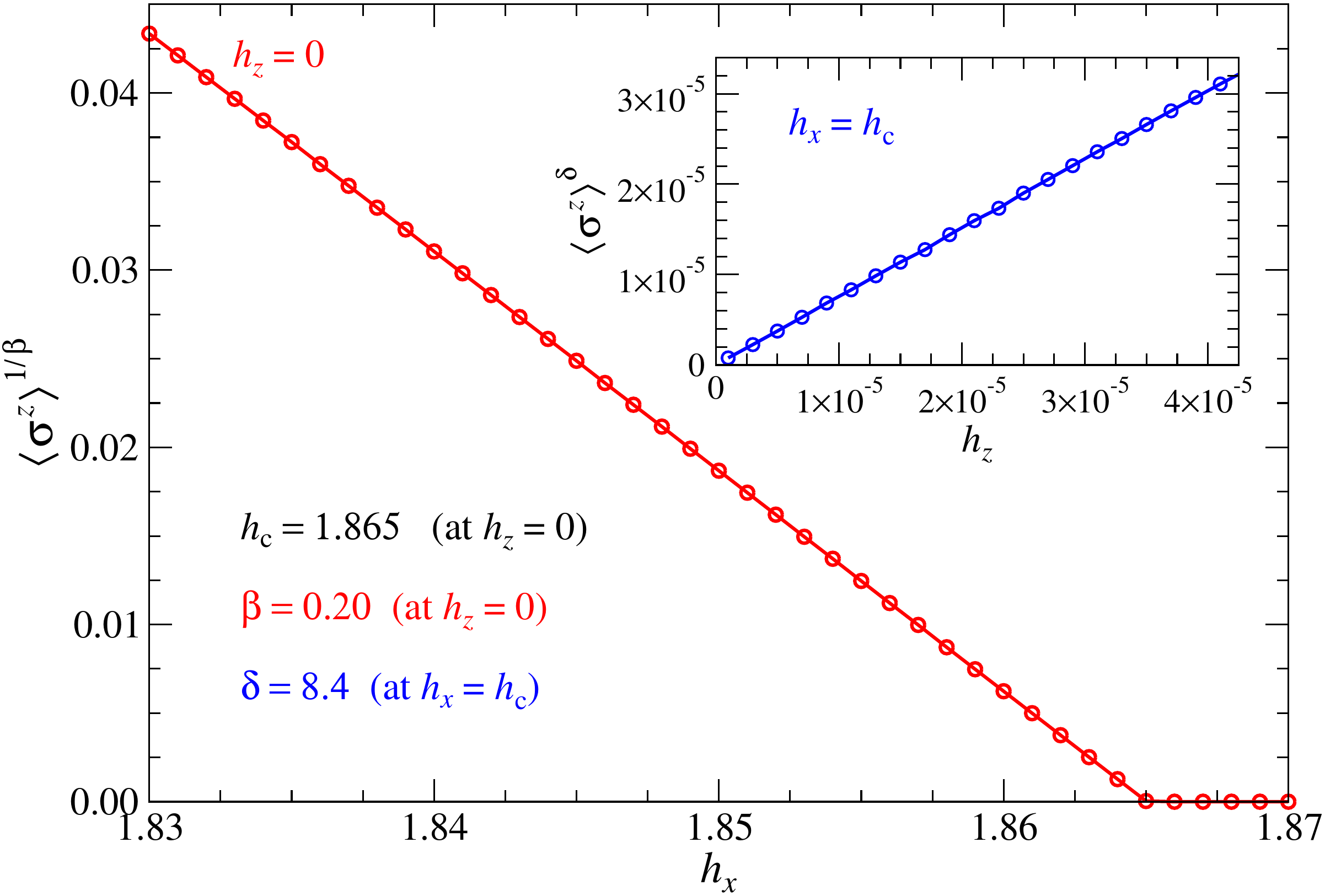}
\caption{(Color online) Linear behavior in $\langle \sigma^z\rangle^{1/0.20}$ below $h_{\rm c}^{~}$, 
which means $\beta = 0.20$. The inset shows linear behavior in $\langle \sigma^z\rangle^{8.7}$ 
with respect to the longitudinal field $h_z^{~}\to 0$ at the criticality $h_x^{~} = h_{\rm c}^{~}$;
the linear behavior occurs at $\delta=8.7$. These values are obtained by the HOTRG method at $D = 20$.}
\label{fig:zmag}
\end{figure}

Finally, we increase $D$ in the HOTRG method to $D = 20$, which is still computationally feasible,
in order to determine the critical field $h_{\rm c}^{~}$ and the critical exponents $\beta$ and $\delta$ 
precisely. The exponent $\beta$ is associated with the critical behavior of the spontaneous magnetization 
$\langle \sigma^z_{~} \rangle \propto ( h_{\rm c}^{~} - h_x^{~} )^{\beta}_{~}$. 
Assuming the scaling form and applying the least-square fitting  to the calculated $\langle \sigma^z\rangle$,
we obtain $h_{\rm c}^{~} = 1.865$ and $\beta = 0.20$. For confirmation, we plot
$\langle \sigma^z_{~} \rangle^{1/\beta}_{~}$ with $\beta = 0.20$ in Fig.~\ref{fig:zmag},
where the linear behavior below $h_{\rm c}^{~}$ is evident. The other exponent $\delta$ 
is associated with the scaling $\langle \sigma^z_{~} \rangle \propto h_z^{1/\delta}$ 
at the critical field $h_x^{~} = h_{\rm c}^{~}$. In the inset, we show 
$\langle \sigma^z_{~} \rangle^{\delta}$ with respect to $h_z^{~}$, which is linear if we assume 
$\delta = 8.7$.

\section{Summary}

The transverse-field Ising model on the Sierpi\'nski fractal was studied by the three methods: (1) the mean-field
approximation, (2) the RSRG method, which can be easily adapted for the fractal structure, and (3) the HOTRG 
method, which had reproduced very reliable results for the transverse-field Ising model on the square lattice~\cite{hotrg}. 
The numerical algorithm in the original HOTRG method has been generalized in order to contract a tensor network
with the fractal structure. We performed the entanglement-entropy analysis in the HOTRG method at the initial stage, 
in order to stabilize the numerical calculation, as shown in the Appendix.

We have confirmed the existence of the second-order phase transition in the quantum Ising model on the 
Sierpi\'nski fractal, whose Hausdorff dimension $d_{\rm H}^{~} \approx1.585$. The critical field is 
$h_{\rm c}^{~} = 1.865$, and the two critical exponents $\beta = 0.20$ and $\delta = 8.7$ are obtained.
Our results are in a good agreement with the MC simulations by Yi~\cite{Yi}, which resulted in
$h_{\rm c}^{~} = 1.865(2)$ and $\beta = 0.19(2)$. Table~\ref{tab1} summarizes the transition point
$h_{\rm c}^{~}$ and the exponents $\beta$ and $\delta$ for the transverse-field Ising model on 
the chain, the Sierpi\'nski fractal, and the square lattice.

\begin{table}[tb]
\tabcolsep=6pt
\caption{Comparison of $h_{\rm c}^{~}$, $\beta$, and $\delta$ for the transverse-field Ising model on 
the chain ($d_{\rm H}^{~} = 1$), the Sierpi\'nski fractal ($d_{\rm H}^{~} = \log_2 3$)~\cite{Yi},
and the square lattice ($d_{\rm H}^{~} = 2$)~\cite{hotrg,cam}.}
\begin{center}
\begin{tabular}{|c|c|c|c|c|}
\hline
$d_{\rm H}$ & $h_{\rm c}^{~}$ & $\beta$ & $\delta$ & method used \\
\hline
$\log_2 2 = 1 \hfill$    & 1\phantom{.0000} & 0.125\phantom{0} & 15  & exact solution \\
\hline
$\log_2 3 \approx 1.585$ & 1.865\phantom{0} & 0.20\phantom{00} & 8.7 & MC, HOTRG       \\
\hline
$\log_2 4 = 2 \hfill$    & 3.0439           & 0.3295           & 4.8 &  CAM, HOTRG    \\
\hline
\end{tabular}
\end{center}
\label{tab1}
\end{table}

Relation between critical behavior and the fractional dimensionality has not been fully investigated yet, and
there are open problems to be considered. One of them is the classification of the quantum phase transition on
the fractal lattice with $d_{\rm H}^{~} = \log_4 12$ that was recently studied for the classical
system~\cite{jozofraktal}. This particular fractal can be dealt with the HOTRG method, as we have considered. 
It is also possible to generate a set of fractal lattices, which have the Hausdorff dimensions $1 < d_{\rm H} < 2$, 
as extensions. How does the the hyper-scaling relations look like on fractal lattices?

Recent studies on neural networks~\cite{qbm,lloyd} have some aspects in common with the current study,
in the point that the formation of complex network geometry is required. 
Investigations of the quantum phase transitions on such non-typical lattices 
could be of use for the initial parameterization of the neural networks with complex geometry.
So far, in the field of tensor-network, supervised machine learning~\cite{SML} 
and quantum machine learning~\cite{QML} were performed on regular lattices.
We conjecture that tensor networks with fractal structure, such as the tree tensor network, 
could lead to an efficient approximation of a given probability distribution in machine learning.

\begin{acknowledgments}
This work was supported by the projects EXSES APVV-16-0186 and VEGA Grant No. 2/0130/15.
T.~N. and A.~G. acknowledge the support of Grant-in-Aid for Scientific Research.
J.~G. and T.~N. were supported by JSPS KAKENHI Grant Number 17K05578 and P17750.
\end{acknowledgments}

\appendix*
\section{Remarks on initialization}

It is known that the Trotter-Suzuki decomposition in Eq.~\eqref{TSD} introduces an error of the order of
$( \Delta \tau )^2_{~}$. Thus it is more suitable to keep $\Delta \tau$ relatively small. We choose
$\Delta \tau$ of the order $10^{-2}$ in most of the numerical calculations. When $\Delta \tau$ is
too small, however, there is a conspicuous anisotropy between the space and imaginary-time directions.
A naive application of the iterative processes in Eqs.~\eqref{HOTRGa}-\eqref{HOTRGc} can cause
numerical instabilities, especially, near the critical field $h_c$. 
This occurs when $\Delta \tau$ is very small, because the local tensor $T^{(0)}_{~}$ works as
an identity in the imaginary time direction, and this situation requires to keep huge $D$ in the
vertical direction, which is not feasible for realistic computational resources.

In order to avoid the instability, we modify the definition of the initial 
tensor $T_{ijk,st}^{(0)}$ by implementing the ``vertical stacking'' of the original local tensor, 
before we start the main iterations in Eqs.~\eqref{HOTRGa}-\eqref{HOTRGc}. 
As we explain here, the succeeding stacking process, which is followed by the renormalization-group
transformation, gradually makes the local tensor isotropic. We introduce two quantities, the planer 
and the vertical entanglement entropies. Identifying all the tensor element as a kind of quantum wave
function, it is understood that the relation between these two entanglement entropies quantifies the
anisotropy in the initial tensor. 

\begin{figure}[tb]
\includegraphics[width = 0.48\textwidth,clip]{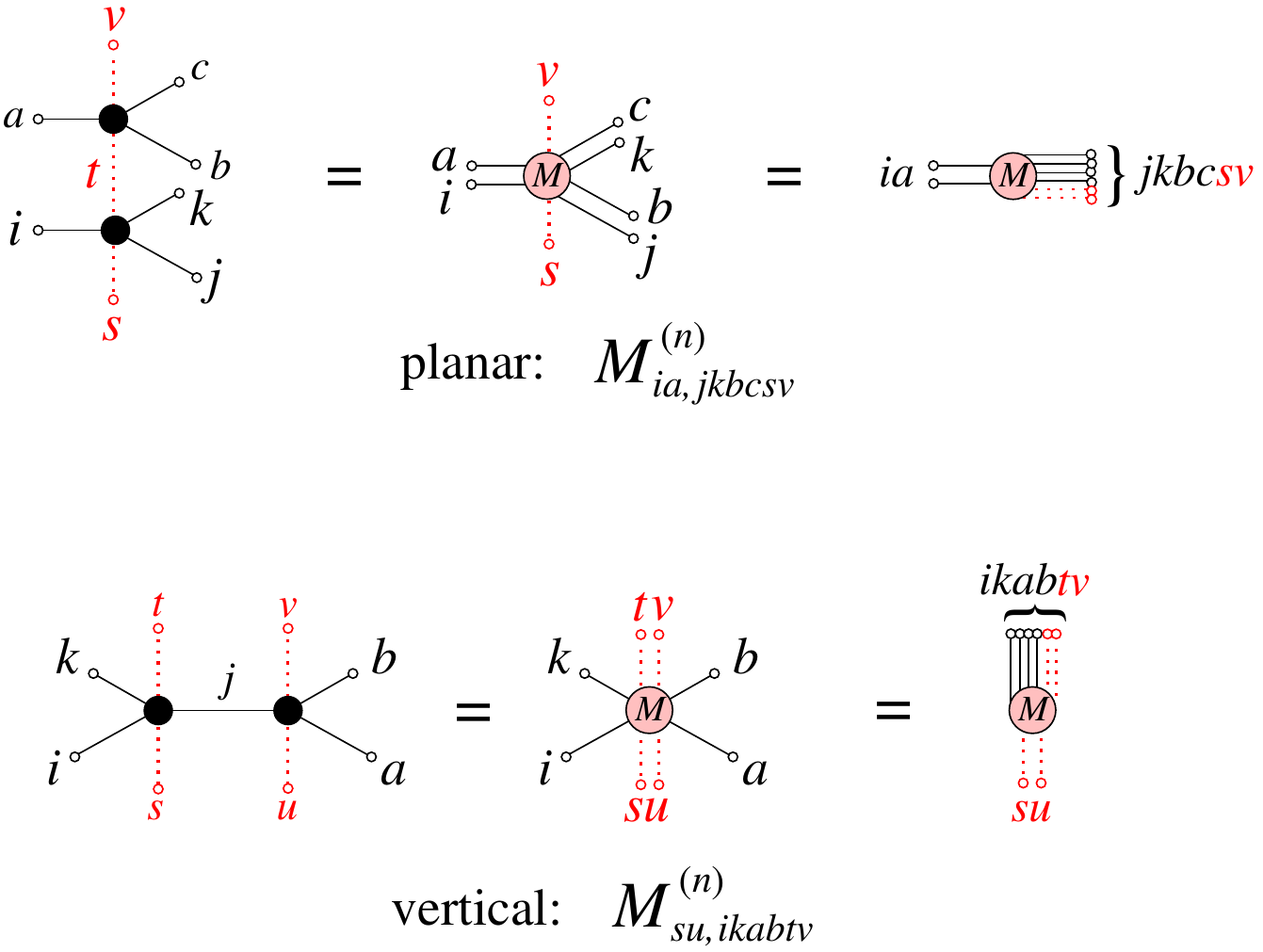}
\caption{(Color online) Graphical representations of Eqs.~\eqref{Mplan} and \eqref{Mplan2}
with the planar (top) and the vertical (bottom) constructions of the $M$ tensors,
respectively. These tensors can be interpreted as rectangular matrices of the size $D^2\times2^2D^4$ (top)
and $2^2\times2^2D^4$ (bottom), respectively, and the singular value decomposition in
Eqs.~\eqref{eqSVD} and \eqref{eqSVD2} is applied.}
\label{fig:Mtensor}
\end{figure}

Let us introduce a new notation $T^{(0,n)}_{~}$, where the integer $n = 0,1,2,\dots$ enumerates the number
of the stacking processes. The first one $T^{(0,n=0)}_{~}$ is the original local tensor, whose element is the 
$T_{ijk,st}^{(0)}$ in Eq.~\eqref{iniT}. The tensor $T^{(0,n+1)}_{~}$ is obtained recursively by stacking two
$T^{(0,n)}_{~}$ vertically and performing the contraction
\begin{equation}
T^{(0,n+1)}_{ijk,st} = \sum_{\genfrac{}{}{0pt}{}{abcd}{efu}}^{~}
T^{(0,n)}_{abc,su} \, T^{(0,n)}_{def,ut} \, U^{(n)}_{ad,i} \, U^{(n)}_{be,j} \, ,
U^{(n)}_{cf,k} \, ,
\label{PI}
\end{equation}
which is essentially the same as Eq.~\eqref{HOTRGc}. Figure \ref{fig:Gtrg}(c) shows the graphical representation
of this process. The isometry $U^{(n)}_{~}$ is obtained as follows. We first combine the two identical tensors 
$T^{(0,n)}_{~}$ {\it vertically} 
\begin{equation}
M^{(n)}_{ia,jkbcsv} = \sum_{t=1}^{D} T^{(0,n)}_{ijk,st} \, T^{(0,n)}_{abc,tv}\, ,
\label{Mplan}
\end{equation}
as shown in the graphical representation in Fig.~\ref{fig:Mtensor} (top). We perform the singular value
decomposition~\cite{hosvd} (SVD) that factorizes $M^{(n)}_{~}$ as 
\begin{equation}
M^{(n)}_{ia,jkbsv} = 
\sum_{\xi=1}^{D^2_{~}}
U^{(n)}_{ia,\xi} \,\, \omega^{(n)}_{\xi} \, V^{(n)}_{\xi,jkbsv}\, .
\label{eqSVD}
\end{equation}
This SVD specifies the isometry $U^{(n)}_{~}$ we need in Eq.~\eqref{PI}. It should be noted that 
the singular values $\omega^{(n)}_{\xi} \geq 0$ plays an important role in both the renormalization-group
transformation and the determination of the entanglement entropy. In the contraction with $U^{(n)}_{~}$,
we keep $D$ largest singular values from $D^2_{~}$ ones, and discard the rest of them. 
One finds that $T^{(0,n)}_{~}$ corresponds to the stack of $2^n_{~}$ numbers of $T^{(0,0)}_{~}$, 
which is contracted by the tree-tensor network constructed by $U^{(m)}_{~}$ for $m = 0$ up to $m = n-1$. 

Let us identify $M^{(n)}_{ia,jkbsv}$ in Eq.~(A.2) as a kind of another quantum wave function
$\Psi_{ia,jkbsv}^{~}$ in order to define the {\it planar} entanglement entropy
\begin{equation}
\varepsilon^{(n)}_{\rm planar} = -\sum_{\xi=1}^{D^2}
\frac{\left[ \omega^{(n)}_{\xi} \right]^2_{~}}{\Omega_n}  \ln  
\frac{\left[ \omega^{(n)}_{\xi} \right]^2_{~}}{\Omega_n}
\end{equation}
for the division of the index into $ia$ and $jkbsv$, where we have used the normalization
\begin{equation}
\Omega_n = \sum_{\xi=1}^{D^2} \left[ \omega^{(n)}_{\xi} \right]^2_{~} 
\end{equation}
for the probability. The entanglement entropy $\varepsilon^{(n)}_{\rm planar}$ quantifies how strongly
the part of the `quantum' system, specified by the indices $ia$, is correlated with the rest of the system,
as specified by the indices $jkbsv$, cf Eq.~\eqref{eqSVD}. (It should be noted that the {\it planer}
entanglement entropy $\varepsilon^{(n)}_{\rm planar}$ is obtained after stacking $T^{(0,n)}_{~}$
{\it vertically}.)

A way of quantifying the anisotropy in $T^{(0,n)}_{~}$ is to observe the entanglement entropy
in the {\it vertical} direction.  Figure~\ref{fig:Mtensor} (bottom) shows the {\it horizontal}
contraction between the two $T^{(0,n)}_{~}$
\begin{equation}
{\tilde M}^{(n)}_{su,ikabtv}  = \sum_{j}^{~} \, T^{(0,n)}_{ijk,st} \, T^{(0,n)}_{jab,uv} \, .
\label{Mplan2}
\end{equation}
Performing the singular value decomposition,
\begin{equation}
{\tilde M}^{(n)}_{su,ikabtv} = 
\sum_{\xi=1}^{2^2_{~}}
{\tilde U}^{(n)}_{su,\xi} \, \, {\tilde \omega}^{(n)}_{\xi} \, {\tilde V}^{(n)}_{\xi,ikabtv} \, ,
\label{eqSVD2}
\end{equation}
we obtain the singular values ${\tilde \omega}^{(n)}_{\xi}$. 
Identifying ${\tilde M}^{(n)}_{su,ikabtv}$ as a kind of quantum wave function $\Phi_{su,ikabtv}^{~}$, 
we obtain the {\it vertical} entanglement entropy
\begin{equation}
\varepsilon^{(n)}_{\rm vertical} = -\sum_{\xi=1}^{2^2_{~}}
\frac{\left[ {\tilde\omega}^{(n)}_{\xi} \right]^2_{~}}{\tilde \Omega_n} \, \ln \, 
\frac{\left[ {\tilde\omega}^{(n)}_{\xi} \right]^2_{~}}{\tilde \Omega_n} \, , 
\end{equation}
where we have used the normalization
\begin{equation}
{\tilde\Omega}_n = \sum_{\xi=1}^{2^2} \left[ {\tilde\omega}^{(n)}_{\xi} \right]^2_{~} \, .
\end{equation}
This {\it vertical} entanglement entropy $\varepsilon^{(n)}_{\rm vertical}$ quantifies the quantum
correlations carried by the indices $su$ in the vertical direction.  

\begin{figure}[tb]
\includegraphics[width = 0.48\textwidth,clip]{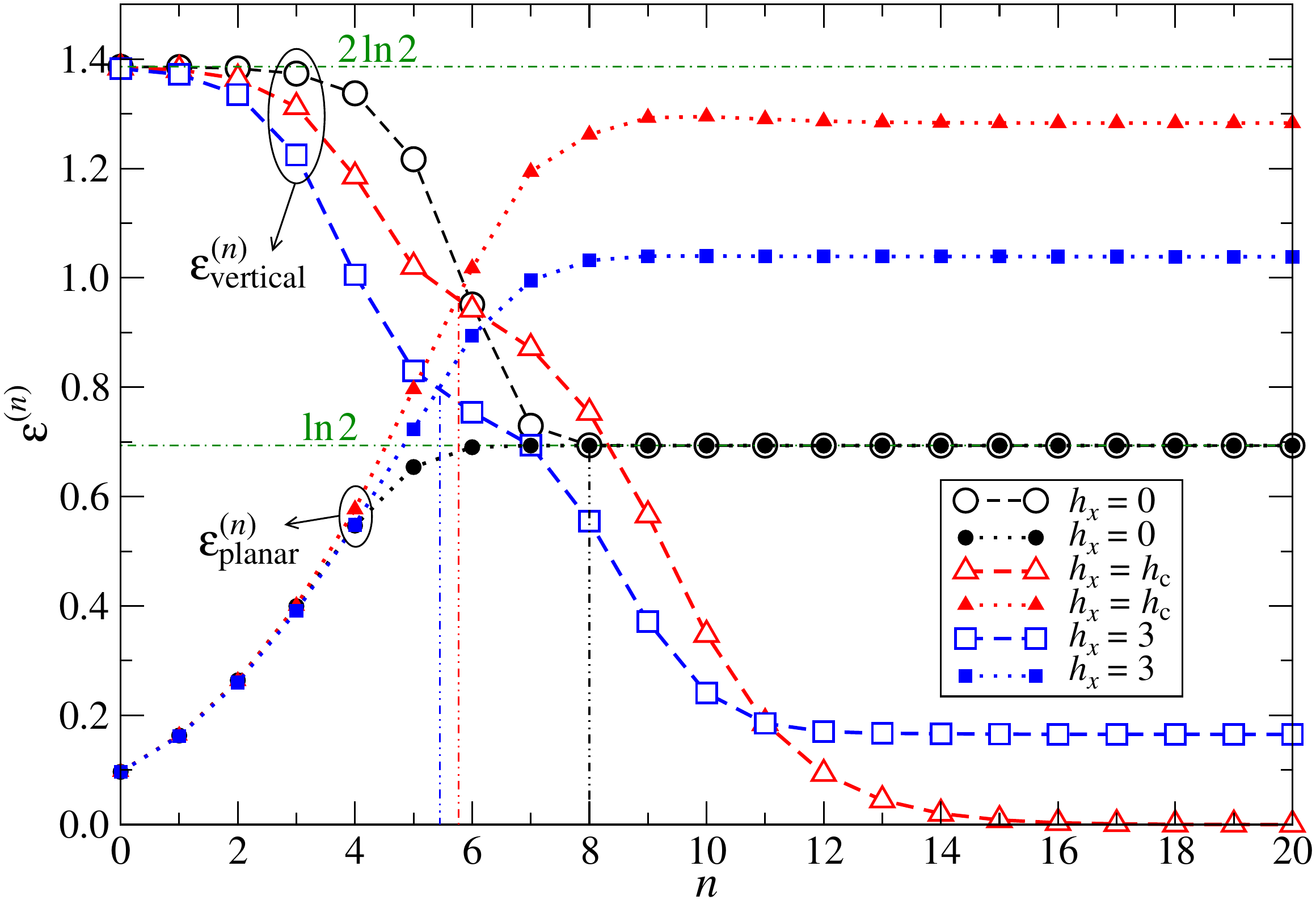}
\caption{(Color online) The entanglement entropy $\varepsilon^{(n)}_{\rm planar}$ and 
$\varepsilon^{(n)}_{\rm vertical}$ with respect to the number of stacking $n$, 
at $h_x^{~} = 0$ (circles), $h_x^{~} = h_{\rm c}^{~}$ (triangles), and $h_x^{~} = 3$ (squares)
for $D = 2$, when $\Delta \tau=0.01$ and $h_z^{~} = 0$.}
\label{fig:entropy}
\end{figure}

We have thus defined $\varepsilon^{(n)}_{\rm planar}$ and $\varepsilon^{(n)}_{\rm vertical}$.
Fig.~\ref{fig:entropy} shows them with respect to $n$ at $h_x^{~} = 0$, $h_x^{~} = h_{\rm c}^{~}$,
and $h_x^{~} = 3$. The tensor $T^{(0,0)}_{~}$ works almost as the identity which is applied to the
vertical direction, and there is almost no correlation to the planar direction. For this reason,
$\varepsilon^{(n)}_{\rm vertical}$ is close to $2 \ln 2$, which corresponds to two completely
entangled pairs, and $\varepsilon^{(n)}_{\rm planar}$ is very small. With increasing $n$,
$\varepsilon^{(n)}_{\rm vertical}$ always decreases, while the planar one $\varepsilon^{(n)}_{\rm planar}$
increases. For $h_x^{~}=0$, both of the entropies are saturated to $\ln D = \ln 2$ at larger $n$
since the calculations are carried out with $D = 2$.

When $\varepsilon^{(n)}_{\rm planar}$ and $\varepsilon^{(n)}_{\rm vertical}$ are close, it is possible
to consider that $T^{(0,n)}_{~}$ is almost equally correlated with both the planar and the vertical
directions, and this is the right situation to start the iterative processes in the HOTRG method.
The vertical double-dot-dashed lines in Fig.~\ref{fig:entropy} shows the value of such $n$. 
Within the typical range of the transverse field $0 \leq h_x^{~} \leq 3$ we have used, the optimal
number of the initial stacking lies in the interval $5 \lesssim n \lesssim 8$. We have numerically
confirmed that the HOTRG method, following Eqs.~\eqref{HOTRGa}-\eqref{HOTRGc}, can be performed in
a stable manner after they have been started with $T^{(0,n\approx6)}_{~}$.

\begin{figure}[tb]
\includegraphics[width = 0.48\textwidth,clip]{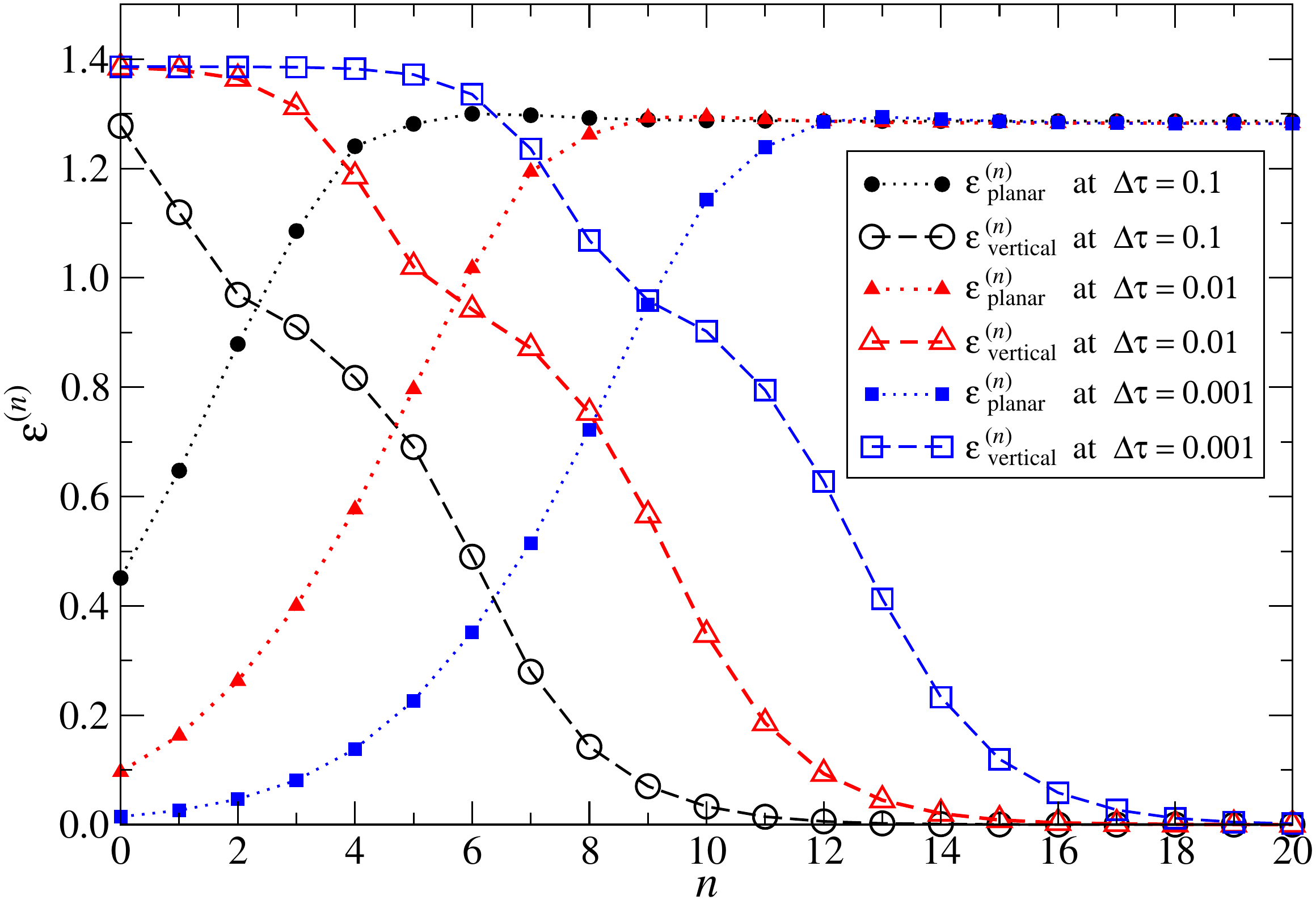}
\caption{(Color online) The 
entanglement entropies
at $h_x^{~} = h_{\rm c}^{~}$ for the three selected imaginary-time
steps $\Delta\tau=0.1$ (circles), $\Delta\tau=0.01$ (triangles), and $\Delta\tau=0.001$ (squares).}
\label{fig:entropy2}
\end{figure}

The correct determination of $n$ also depends on the initial choice of $\Delta \tau$. Figure~\ref{fig:entropy2} 
shows both of the entropies at the critical field $h_x^{~} = h_{\rm c}^{~}$ for three selected imaginary-time 
steps $\Delta \tau$. As it is naturally understood, the smaller the $\Delta \tau$, the more initial iteration
steps $n$ are necessary to satisfy $\varepsilon^{(n)}_{\rm planar} \approx \varepsilon^{(n)}_{\rm vertical}$.
Although smaller $\Delta \tau$ lowers the Trotter-Suzuki error $(\Delta \tau)^2$, significantly more
iterations are needed in the main HOTRG algorithm and round-off errors get accumulated for
$\Delta \tau \ll 10^{-2}$, which negatively act against the improvement of numerical precision.
Thus we have used $\Delta\tau$ of the order of $0.01$ for all the calculations in the main text.

\bibliography{sierpinski.bib}

\end{document}